\documentclass[aps,prd,showpacs,twocolumn,superscriptaddress,preprintnumbers,10pt,floatfix]{revtex4}

\usepackage{amssymb}  % \gtrsim, \geqslant, etc etc: see amsguide.ps
\usepackage{graphicx}

\usepackage[dvips]{color}

\newcommand{\al}{\alpha}

\newcommand{\de}{\delta}
\newcommand{\De}{\Delta}
\newcommand{\ep}{\varepsilon}

\newcommand{\la}{\lambda}

\newcommand{\si}{\sigma}

\newcommand{\beq}{\begin{equation}}
\newcommand{\eeq}{\end{equation}}
\newcommand{\ba}{\begin{array}}
\newcommand{\ea}{\end{array}}
\newcommand{\bea}{\begin{eqnarray}}
\newcommand{\eea}{\end{eqnarray}}
\newcommand{\bi}{\begin{itemize}}  %\setlength{\itemsep}{0\parsep}}
\newcommand{\ei}{\end{itemize}}
\newcommand{\ben}{\begin{enumerate}} %\setlength{\itemsep}{0\parsep}}
\newcommand{\een}{\end{enumerate}}
\newcommand{\bc}{\begin{center}}
\newcommand{\ec}{\end{center}}

\newcommand{\p}{\partial}
 
\newcommand{\txt}{\textstyle}
\newcommand{\dsp}{\displaystyle}
\newcommand\eqn[1]{(\ref{#1})}      % parentheses around the LaTex "ref" macro
\newcommand{\half} {{\txt \frac{1}{2}}}

\newcommand{\fm}{{\rm fm}}

\newcommand{\MeV}{{\rm MeV}}

\newcommand{\mucrit}{\mu_{\rm crit}} 
\newcommand{\sicrit}{\si_{\rm crit}} 
 
\newcommand{\sibar}{{\overline\si}} 
\newcommand{\sibarcrit}{{\overline\si_{\rm crit}}} 
 
\newcommand{\dGbar}{{\overline{\De g}}}

\usepackage[normalem]{ulem}  % \sout{old text} for strikeout

\begin{document}
\title{The Stability of Strange Star Crusts and Strangelets}
\author{Mark G. Alford}
\affiliation{Department of Physics, Washington University, 
  St Louis, MO 63130, USA}
\author{Krishna Rajagopal}
\affiliation{Center for Theoretical Physics, Massachusetts Institute
of Technology, Cambridge, MA 02139, USA}
\affiliation{Nuclear Science Division, MS 70R319,
Lawrence Berkeley National Laboratory, Berkeley, CA 94720, USA}
\author{Sanjay Reddy}
\affiliation{Theoretical Division,
  Los Alamos National Laboratory,  Los Alamos, NM 87545, USA}
  \author{Andrew W. Steiner}
\affiliation{Theoretical Division,
  Los Alamos National Laboratory,
  Los Alamos, NM 87545, USA}

\preprint{MIT-CTP-3731}

\begin{abstract}
We construct strangelets, 
taking into account electrostatic effects, including
Debye screening, and arbitrary surface tension $\si$
of the interface between vacuum and quark matter.
We find that there is a critical surface tension $\si_{\rm crit}$
below which large strangelets are unstable to fragmentation and
below which quark star surfaces will fragment into a crystalline
crust made of charged strangelets immersed in an electron gas.
We derive a model-independent relationship between $\sicrit$
and two parameters that characterize any quark matter equation of state.
For reasonable model equations of state, we find $\sicrit$ typically
of order a few MeV$/$fm$^2$. If $\si\leqslant \sicrit$, the size-distribution
of strangelets in cosmic rays could feature a peak corresponding to
the stable strangelets that we construct.
\end{abstract}


\maketitle

\section{Introduction}
\label{sec:intro}
Ordinary matter consists of atoms, whose nuclei are ultimately
composed of up and down quarks. One can think of nuclei as droplets
of nuclear matter, which is observed to be very stable:
the most stable nuclei have lifetimes longer than the age of the universe.
However, it has been hypothesized 
\cite{Bodmer:1971we,Witten:1984rs,Farhi:1984qu}
that nuclear matter may actually be metastable, and the
true ground state of matter consists of a combination of
up, down, and strange quarks known as ``strange matter''.
Small nuggets of such matter are called ``strangelets''.
If this ``strange matter hypothesis'' is true, then
many if not all compact stars are not neutron stars but ``strange stars'':
large balls of strange matter (for a review see Ref.~\cite{Weber:2004kj}).
Moreover, strangelets may be produced
in compact star collisions, and hence contribute to the cosmic ray
background, or in heavy-ion collisions, where they could be observed
in terrestrial experiments.

Until recently, it has been assumed that the boundary between
strange matter and the vacuum is a simple surface, with a 
layer of positively charged quark matter beneath it and with
electrons floating above it,
sustained by an electric
field which could also support a small normal nuclear crust in
suspension above the quark matter~\cite{Alcock:1986hz}, as long
as the strange star is not too hot~\cite{Usov:1997eg}.

This was recently questioned by Jaikumar, Reddy, and
Steiner~\cite{Jaikumar:2005ne}, who showed that if Debye screening and
surface tension were neglected then the surface must actually fragment
into a charge-separated mixture, involving positively-charged
strangelets immersed in a negatively charged sea of electrons,
presumably forming a crystalline solid crust.  At the surface of this
strangelet$+$electron crust, there would be no electric field.  This
gives a radically different picture of the strange star surface,
making it much more similar to that of an ordinary neutron star, and
casts doubt on all of the previous work on the phenomenology of
strange star surfaces.

In this paper we address the stability of electrically neutral bulk
quark matter with respect to fragmentation into a charge-separated
strangelet$+$electron crust, and the almost precisely equivalent
question of whether
large strangelets are stable with respect to fission into smaller strangelets.
We shall include energy costs due to Debye screening, neglected
in Ref.~\cite{Jaikumar:2005ne}, and surface tension.  Our treatment
of Debye screening follows that of Heiselberg~\cite{Heiselberg:1993dc} except
that we include the energy gained by charge separation which was neglected
in Ref.~\cite{Heiselberg:1993dc}  
and whose importance was discovered in Ref.~\cite{Jaikumar:2005ne}.

We identify the three microscopic parameters that play a crucial role
in the stability analysis: (1) the charge density of quark matter at
zero electric charge chemical potential denoted as $n_Q$; (2) the
electric charge susceptibility of quark matter denoted as $\chi_Q$;
and (3) the surface tension between quark matter and vacuum, denoted
as $\sigma$.  In Sections \ref{sec:stability} and
\ref{sec:construction} we derive all of our results in terms of these
parameters. In Section \ref{sec:stability} we give a parametric
estimate of the critical surface tension below which neutral quark
matter and large strangelets are unstable to fragmentation, and in
Section \ref{sec:construction} we construct explicit strangelet
profiles and thus determine the dimensionless quantities left
unspecified in Section \ref{sec:stability}.

The quantities $n_Q$ and $\chi_Q$ can be evaluated for any proposed
quark matter equation of state, and in Section
\ref{sec:model-dependent} we shall do so in bag models for unpaired
quark matter and quark matter in the 2SC color superconducting
state~\cite{Bailin:1983bm,Alford:1997zt,Rapp:1997zu,Reviews}.  We find
that even if the strange quark is so heavy that no BCS pairing occurs
in bulk quark matter, if such bulk matter is unstable to fission then
the strangelets that result are likely to be in the 2SC phase.
This occurs because charged 2SC matter may have higher pressure
than charged unpaired quark matter, even when neutral 2SC matter has
lower pressure than neutral unpaired quark matter.

The color-flavor locked (CFL)
phase~\cite{Alford:1998mk,Reviews} is the favored phase
of neutral quark matter if the density is high enough, the pairing
interactions are strong enough, or the strange quark is light enough.
The CFL phase is an insulator, so both $n_Q$ and
$\chi_Q$ vanish, because there is pairing between all three flavors,
and no electrons are needed in order to maintain
neutrality.  This makes CFL matter stable against fission even if
$\sigma$ is arbitrarily small. Our goal in this paper is to understand
what happens if the
quark matter at the surface of a strange star is not in the CFL phase,
although this form of quark matter may occur deeper within.

We close in Section \ref{sec:discussion} with comments on
phenomenological implications for the strange star surface and
experimental strangelet searches.

\section{Stability of Strangelets}
\label{sec:stability}

The stability of strangelets is determined by the equation of state
(EoS) for quark matter (QM), which will be discussed below,
and the EoS of the vacuum with electrons, which is well known.
%$(p_\vac(\mu,\mu_e) =\frac{\mu_e^4}{12\pi^2} for massless electrons)
The equation of state expresses the pressure $p(\mu,\mu_e)$ as a 
function of the chemical
potentials for the two conserved quantities,
quark number ($\mu$) and negative electric charge ($\mu_e$).
We work at zero temperature throughout.
The quark density $n$ and the electric charge density $q$ (in units
of the positron charge $e$) are
\beq
%\ba{rcl}
n = \frac{\p p}{\p \mu},\quad 
q = \frac{\rho}{e} = -\frac{\p p}{\p \mu_e} \ .
%\ea
\label{charges}
\eeq
The chemical potential $\mu_e$ is the energy cost of
adding an electron, so it is the positron charge $e$ multiplied by
the electrostatic potential $\phi$. Hence, $\mu_e q = \phi \rho$.
The energy densities follow from the usual thermodynamic relation
\beq
\ba{rcl}
\ep = \mu n - \mu_e q - p\ .
\ea
\label{energy_density_v1}
\eeq
The second term in \eqn{energy_density_v1} and the final term in
\eqn{charges} are negative because $\mu_e$ is the chemical potential
for {\em negative} electric charge. As written, $\ep$ and $p$ are understood
to include the electric
field energy 
and pressure.  Henceforth, however, we shall intepret $p$ as only the
kinetic pressure of the quark matter and electrons, in which case the electric field energy
$\half \int d^3 r \phi\rho$ must be added to $\int d^3 r \varepsilon$. This
can be implemented by writing
\beq
\ba{rcl}
\ep = \mu n - \half \mu_e q - p_{\rm QM + e}\ .
\ea
\label{energy_density}
\eeq

We make the strange quark matter hypothesis that 
at some $\mu_{\rm crit}$, uniform neutral quark matter with some
nonzero $n$ has the same pressure as the vacuum.
(The hypothesis further requires that $\mucrit < \mu_{\rm nuclear}$,
 where $\mu_{\rm nuclear}\approx 310~\MeV$ is the chemical potential of
uniform neutral nuclear matter at zero pressure, and that
the chemical potential for two-flavor quark matter with zero
pressure is greater than $\mu_{\rm nuclear}$.)   However, if only
volume contributions to the energy are taken into
account, then  low pressure
quark matter is unstable against fragmentation into positively
charged strangelets embedded in a gas of electrons~\cite{Jaikumar:2005ne}.
In this paper we evaluate the surface tension required to stabilize
neutral quark matter at zero pressure against such fragmentation,
taking into account Debye screening.

Debye screening causes the positive charge density in large
strangelets to migrate towards the surface, resulting in a charged
skin whose thickness is of order the Debye length $\la_D$ and a
neutral interior.  So, for large strangelets ($R\gg \la_D$) the volume
energy benefit of charge separation is reduced to a surface energy.
Typically, $\la_D$ is of order 5~fm in quark matter.  Whereas Debye
screening penalizes large strangelets, surface tension penalizes small
strangelets most, as their surface to volume ratio is the greatest.
Both effects make fragmentation into positively charged strangelets
less favorable. We can expect their combined effect
to be least for droplets with $R\sim\la_D$, meaning that if bulk quark
matter proves unstable, it will fragment into strangelets of this
size.

Jaikumar et al.~proposed that the outer layers of a strange star could
be constructed from positively charged strangelets with some radius
$R$ embedded in a gas of electrons, forming a Wigner-Seitz
lattice~\cite{Jaikumar:2005ne}.  The local pressure $p$ will vary
considerably within a single Wigner-Seitz cell, but the external
pressure $p_{\rm ext}$ on a Wigner-Seitz cell from its neighbours will
be zero at the surface of the star, and increase smoothly with depth.
To evaluate the stability of such a crust, we must evaluate the Gibbs
free energy per quark
\beq
g(R) = \frac{E(R) + p_{\rm ext}V(R)}{N(R)}\ , 
\eeq
where $E$, $V$ and $N$ are the energy, volume
and quark number of a Wigner-Seitz cell containing a strangelet
of radius $R$.
We shall calculate the energy $E(R)$ by
solving the Poisson equation upon making a Thomas-Fermi approximation
to obtain the distribution of charge, integrating $\ep$ as defined in
\eqn{energy_density}, and then adding the surface tension energy
$4\pi R^2\si$.  We must then compare $g(R)$ with the 
Gibbs free energy per quark $g_{QM}$ of
uniform neutral quark matter at the same pressure $p_{\rm ext}$.  We
shall work throughout at $p_{\rm ext}=0$, meaning that we analyze the
stability of the outer surface of the mixed phase crust proposed in
Ref.~\cite{Jaikumar:2005ne} and at the same time that of an isolated
strangelet with radius $R$.  We shall be comparing them to neutral
bulk quark matter with $\mu=\mucrit$, which has zero pressure and
$g_{QM}=\mucrit$.
So, at $p_{\rm ext}=0$, the free energy per quark
of a strangelet of radius $R$, relative to infinite neutral quark matter,
is 
\beq
\De g(R) = g(R)-g_{QM} = \frac{E(R)}{N(R)} - \mucrit\ .
\label{Delta_g}
\eeq

For any $\si$, $\De g(R)\rightarrow 0$ for $R\rightarrow \infty$.  For
large enough values of $\sigma$, $\Delta g(R)$ will be dominated by
its surface energy contribution $3\sigma/(nR)$, making $\De g(R)>0$
for all $R$.  This means that for large enough $\sigma$ neutral bulk
matter is stable with respect to fragmentation.  For large enough
$\sigma$, furthermore, $\De g(R)$ decreases monotonically with
increasing $R$, meaning that isolated strangelets of any size are
stable, but can lower their energy by fusing with other strangelets
should they encounter them.  At the other extreme, with $\sigma=0$ we
know from the work of Ref.~\cite{Jaikumar:2005ne} that $\De g(R)<0$
for small enough $R$. We therefore expect that for small enough values
of $\sigma$ we shall find a range of $R$ in the vicinity of
$R\sim\lambda_D$ for which $\De g(R)<0$.  This means that, for small
enough $\sigma$, neutral bulk quark matter and large strangelets are
unstable to fragmentation, with the stable strangelets being those
having the size $R=R_*$ that minimizes $\De g(R)$.

The equation of state of quark matter at 
phenomenologically interesting densities and $T\ll \mu$ cannot currently
be calculated by lattice gauge theory or by other methods, so we can
either obtain an approximate EoS from some model, or make a general
parameterization.
We will find that $\mu_e\ll \mu$ for all strangelets
in all models that we consider,
so a general parameterization of the EoS can be obtained by
expanding in powers of $\mu_e/\mu$ \cite{Jaikumar:2005ne},
\begin{equation}
p_{\rm QM}=p_0(\mu,m_s)-n_Q(\mu,m_s)\mu_e 
+ \half\chi_Q(\mu,m_s) \mu_e^2+\ldots
\label{generic_EoS}
\end{equation}
This second-order expansion, which neglects the 
electron pressure $p_e\sim\mu_e^4$, can be used for any model EoS or for
that predicted by QCD. We will see below that it
is an excellent approximation
for the analysis of strangelets. It reduces the EoS-dependence
to specifying the three functions 
$p_0$, $n_Q$ and $\chi_Q$. Moreover, only $n_Q$ and $\chi_Q$
occur in the Poisson equation.

Neutral bulk quark matter has $q=0$, so its electron chemical potential
(\ref{charges}) is
\beq
% \mu_e=
\mu_e^{\rm neutral} \equiv \frac{n_Q}{\chi_Q}\ .
\label{mueneutral}
\eeq
The quark chemical potential $\mucrit$ at which it has zero pressure
is determined by solving 
\beq
p_0(\mucrit,m_s) = \frac{n_Q^2(\mucrit,m_s)}{2\,\chi_Q(\mucrit,m_s)}\ .
\label{p0expression}
\eeq
The Debye screening length in quark matter is controlled by $\chi_Q$, and
is given by
\beq
\lambda_D=\frac{1}{\sqrt{4 \pi \al\chi_Q}}\ ,
\label{debyedefn}
\eeq
with $\alpha=1/137$.

We close this section with a parametric estimate of the critical
surface tension below which strangelets with $R\sim\lambda_D$
have $\Delta g(R)<0$, making neutral bulk quark matter (and larger
strangelets) unstable to fragmentation. 
In section \ref{sec:construction} we will quantitatively determine 
$\mu_e(r)$ and the radius $R_*$ of the most stable strangelet.
For now, we write $x_*\equiv R_*/\lambda_D$ 
and take $\mu_e$ in the strangelet to be a constant with the value 
$f n_Q/\chi_Q$. Presuming the dimensionless quantities
$x_*$ and $f$ to be of order unity, we use
Eqs.~\eqn{charges}, \eqn{energy_density}, \eqn{Delta_g}, \eqn{generic_EoS},
\eqn{mueneutral}, \eqn{p0expression} and \eqn{debyedefn} to obtain
\beq
\De g(R_*) = \frac{3\si}{nR_*} -\frac{n_Q^2(1-f)}{2n\chi_Q }
\label{Delta_g_estimate}
\eeq
which is negative if $\si<\si_{\rm crit}$ where
\beq
\si_{\rm crit} = \frac{(1-f)R_* n_Q^2}{6\chi_Q} 
  = \frac{(1-f)x_*n_Q^2}{12\sqrt{\pi\al}\chi_Q^{3/2}} \ .
\label{crude_estimate}
\eeq
In the next section we shall construct strangelets
by solving the Poisson equation and show that this parametric estimate is
valid, with $x_*\approx 1.61$ and $f\approx 0.49$.

\section{Constructing strangelets}
\label{sec:construction}

We assume that the lowest energy state of a strangelet of radius $R$ is
spherically symmetric. The quark chemical potential $\mu$ is independent
of $r$ because the only net force on a given part of
the strangelet is the electrostatic force. (We assume our strangelets are
small enough that their self-gravity is unimportant, and that color
is screened so strong interactions do not occur across distances
greater than about 1~\fm.)
%This follows from the requirement that the pressure gradient across
%any part of the strangelet, $dp/dr$, be balanced by the electrostatic
%force on it, $-\rho d\phi/d r = - q d\mu_e/dr$.
%\beq
%\ba{rcl}
 %\dsp\frac{d p}{d r} &=&\dsp -\rho \frac{d\phi}{d r} \\[2ex]
 %\frac{\p p}{\p\mu}\frac{d\mu}{d r}
%+\frac{\p p}{\p\mu_e}\frac{\d\mu_e}{d r} =
%\dsp n \frac{\p\mu}{\p r} + \rho \frac{d\phi}{d r} 
 % &=& \dsp -?\rho \frac{d\phi}{d r} \ .
%\ea
%\eeq
The value of $\mu$ inside the droplet
is a little higher than $\mucrit$ because the surface
tension compresses the droplet slightly.
To determine the value of $\mu$, we require the pressure discontinuity
across the surface of the strangelet to be balanced by the surface tension:
\beq
\lim_{\de r\to 0}\bigl( p(R-\de r) - p(R+\de r)\bigr) = \frac{2\si}{R} \ .
\label{pressure_disc}
\eeq
The larger $\si$ becomes, the more compressed the strangelet is, and the
higher the value of $\mu$ inside the strangelet.

To calculate $\mu_e(r)=e\phi(r)$ in the Thomas-Fermi approximation
we solve the Poisson equation,
which in Heaviside-Lorentz units with $\hbar=c=1$
takes the form
\beq
\ba{rcl}
\nabla^2 \phi(r) &=& - \rho(r) \ , \\
{\rm i.e.}\quad \nabla^2 \mu_e(r) &=& -4 \pi \alpha q(r) \ ,
\ea
\label{Poisson}
\eeq
subject to the boundary conditions
\beq
\ba{rcl}
\dsp \lim_{r\to\infty} r\phi(r) &=& Z_\infty\,e \ ,\\[2ex]
\dsp \frac{d \phi}{d r}(r=0) &=& 0 \ , \\
\dsp \frac{d \phi}{d r}(R^+) - \frac{d \phi}{d r}(R^-)&=& 0 \ . \\
\ea
\label{BC}
\eeq
The second and third boundary
conditions follow from the fact that there are no delta-functions of 
localized charge, so $\phi(r)$ and $d\phi/dr$ are
continuous everywhere.   
The first boundary condition states that the net
charge of the strangelet, including any electrons
inside or around it, is $Z_\infty e$.
The Fermi wavelength of the quarks is $2\pi/\mu$ which is
around 4~\fm, so it is  reasonable
to use the Thomas-Fermi approximation to describe the charge
distribution due to quarks inside a strangelet with 
diameter of order $3\lambda_D \sim 15~\fm$, which is the approximate size
of the stable strangelets that we will find.

Strangelets with size $R\sim \lambda_D$ are too small to
have electrons localized within them, 
meaning that the charge
of the strangelet itself is given by that of the quark matter,
\beq
Z=\int_0^R d^3\, r q(r) = \int_0^R d^3r \left(n_Q-\chi_Q \mu_e(r)\right)\  .
\label{Zvalue}
\eeq
To analyze an isolated strangelet, not surrounded by an atom-sized
cloud of electrons, we must find solutions to the Poisson equation
with $Z_\infty=Z$.
To analyze a strangelet located within the Wigner-Seitz cell
of a strange star crust, however, we must include a cloud of electrons around
the strangelet such that the strangelet and electrons together satisfy
the Poisson equation with $Z_\infty=0$.  (Each Wigner-Seitz cell 
contains
one strangelet and $Z$ electrons, ensuring
that the crust is neutral on macroscopic length scales.)
In our numerical work, we have included the electrons outside the
strangelet, taking $q(r)=-\mu_e^3/3\pi^2$ for $r>R$, and found solutions
to the Poisson equation with $Z_\infty=0$. As in atomic physics, the Thomas-Fermi 
description of the cloud of electrons around a strangelet with charge $Z$
becomes more accurate with increasing $Z$.  
Fortunately, it turns out that the electrons do not play an
important role in the stability of the strangelet, so we can ignore
them, setting $q(r)=0$ for $r>R$ and finding solutions to the Poisson
equation with the boundary condition at infinity 
given simply by $Z_\infty=Z$.  We have done all our analytic calculations 
with this boundary condition, and have confirmed numerically that
adding electrons to the system in order to find solutions with $Z_\infty=0$
makes a negligible difference to the energy $E$ and
to $\mu_e(r)$ inside the strangelet.  This simplification arises
because, as mentioned in Section II, 
the electron contribution to the EoS is subleading in $\mu_e/\mu$.

%However, the Fermi wavelength of the electrons is 
%$2\pi/\mu_e$ which is
%typically of order 50~\fm, much larger than 
%the size of typical stable strangelets. This means that
%if electrons accompany the strangelet, they are spread out in a cloud
%around it, and may not be accurately described by a Thomas-Fermi
%approximation.

%This is because, 
%as mentioned in section \ref{sec:stability}, their contribution
%to the EoS is subleading
%in $\mu_e/\mu$. In our analytic calculations we therefore
%ignore the electrons completely. In the numerical calculations
%we include the contribution of a neutralizing electron cloud in and around
%the strangelet (imagining it to be correctly treated by
%Thomas-Fermi), and confirm that this makes a negligible difference.

%In our analytical treatment, then, we set $q(r)=0$ 
%for $r>R$ and
%$q(r)=n_Q-\chi_Q \mu_e(r)$ for $r<R$, so our
%strangelets have net charge, entirely from the quark matter
%\beq
%Z=\int_0^R d^3r \left(n_Q-\chi_Q \mu_e(r)\right)\ ,
%\label{Zvalue}
%\eeq
%and our boundary condition at infinity is $Z_\infty=Z$. 

%As the close agreement between our analytic and numerical calculations shows,
%adding electrons to the system reduces $Z_\infty$, but makes a 
%negligible difference to $\mu_e(r)$ inside the strangelet.

We obtain $E(R)$ and $N(R)$, the quantities which according
to (\ref{Delta_g}) we need in order to determine the stability of a strangelet
with some specified radius $R$, as follows.
For a given $\mu$ we solve \eqn{Poisson} subject to the boundary
conditions \eqn{BC}. We repeat this for different values of $\mu$ until we
find the one that obeys \eqn{pressure_disc}. We then have $\mu$ and
$\mu_e(r)$, so from \eqn{energy_density} and \eqn{charges}
we can obtain the energy density $\ep(r)$ and quark
number density $n(r)$. We integrate these, 
and add in the
surface energy, to obtain the total
energy $E$ and quark number $N$ and hence the energy per quark $E/N$.

The solution to the Poisson equation satisfying the boundary conditions 
(\ref{BC}) with $Z_\infty=Z$ given by (\ref{Zvalue}) is~\cite{Heiselberg:1993dc}
\beq
\ba{r@{\,:\quad}rcl}
r<R& \mu_e(r)&=&\dsp \frac{n_Q}{\chi_Q}
  \left(1 - \frac{\lambda_D}{r}\frac{ \sinh(r/\lambda_D)}{\cosh(R/\lambda_D)}
  \right) \\[3ex]
r>R& \mu_e(r)&=& \dsp \frac{Z\alpha}{r}
\ea
\label{mueprofile}
\eeq
where
\beq
Z=4\pi n_Q \lambda_D^3 \left(x -
\tanh x\right)\ ,
\eeq
where we have defined $x\equiv R/\lambda_D$.
We see that $\mu_e(r)<\mu_e^{\rm neutral}=n_Q/\chi_Q$ throughout the
strangelet, with $\mu_e(r)$ closer to zero for smaller strangelets and
closer to $\mu_e^{\rm neutral}$ for larger ones.

To this point we have not determined $\mu$. (And, recall that both
$n_Q$ and $\chi_Q$ depend on $\mu$.) Because we have no electrons and
hence $p=0$ outside the strangelet, the condition \eqn{pressure_disc}
requires that just inside the surface of the strangelet the pressure
is given by $2\sigma/R$.  Recalling that $n=\partial p/\partial\mu$
and that $p(\mu_{\rm crit})=0$ and using Eqs.~(\ref{generic_EoS}) and
(\ref{mueprofile}) we see that this requires that the quark chemical
potential inside the strangelet (which is constant throughout its
interior) is given by
\beq
\mu=\mu_{\rm crit} - \frac{n_Q^2}{2 n \chi_Q}\frac{\tanh^2 x}{x^2}
 + \frac{2\sigma}{nR}\ . 
\label{mu_internal}
\eeq
The negative term present even when $\sigma=0$ arises because
$\mu_e(R)<\mu_e^{\rm neutral}$, making $p(\mu_{\rm crit},\mu_e(R))>0$.
To achieve $p=0$, as required just inside the edge of the droplet if
$\sigma=0$, $\mu$ must be less than $\mu_{\rm crit}$.  The
$\sigma$-dependent positive term then enforces the pressure
discontinuity (\ref{pressure_disc}) required at nonzero $\sigma$.
(The derivation requires that each of the corrections to $\mu$ are
separately much smaller than $\mu_{\rm crit}$, which is well satisfied
in all results we show.)

\begin{figure}
\includegraphics[scale=0.7]{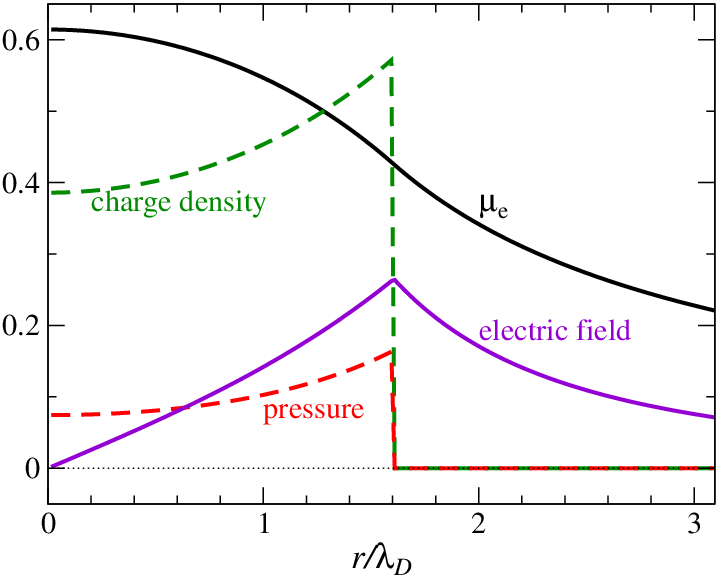}
\caption{
A strangelet profile with radius $R=1.6 \lambda_D$, and
$\sigma=0.1325\, n_Q^2\lambda_D/\chi_Q$. (This corresponds to
the barely-stable strangelet at the critical surface tension---see
Section \ref{sec:construction}.)
The horizontal axis is
$x=r/\lambda_D$. The charge density $q$ is plotted in units of $n_Q$, 
$\mu_e$ is plotted in units of $n_Q/\chi_Q$, 
the electric field is
plotted in units of $n_Q/\sqrt{\chi_Q}$ and the pressure $p_{\rm QM}$
is plotted in units of $n_Q^2/\chi_Q$.  Because of the electric field,
the charge density in the strangelet is pushed towards the outer
edge. The pressure gradient within the strangelet is balanced by the
electric force on the charged matter.
}
\label{fig:profile}
\end{figure}

We can now construct the profiles of $\mu_e(r)$, $q(r)$, the electric
field and $p(r)$ for strangelets with radius $R$. An example is shown
in Fig.~\ref{fig:profile}.  We show the profiles in terms of $n_Q$ and
$\chi_Q$.  Expressed this way, they are model-independent.  Given a
model equation of state, $n_Q$ and $\chi_Q$ must be evaluated at the
$\mu$ given in (\ref{mu_internal}), which itself depends on $n_Q$ and
$\chi_Q$.  This means that the explicit determination of $\mu$ in a
model must be done numerically, although in practice we find that
the critical surface tension is very small, so for $\si\approx\sicrit$
we can use $n_Q(\mucrit)$ and $\chi_Q(\mucrit)$ without making
significant errors.
We have compared the profiles
obtained analytically in terms of $n_Q$ and $\chi_Q$ as we have
described with those obtained numerically, with $n_Q$ and $\chi_Q$
specified by the bag model expressions for unpaired quark matter given
in Section \ref{sec:model-dependent} and with electrons included in
the solution of the Poisson equation. The agreement between analytical
and numerical profiles is excellent.

\begin{figure}
\includegraphics[width=\hsize]{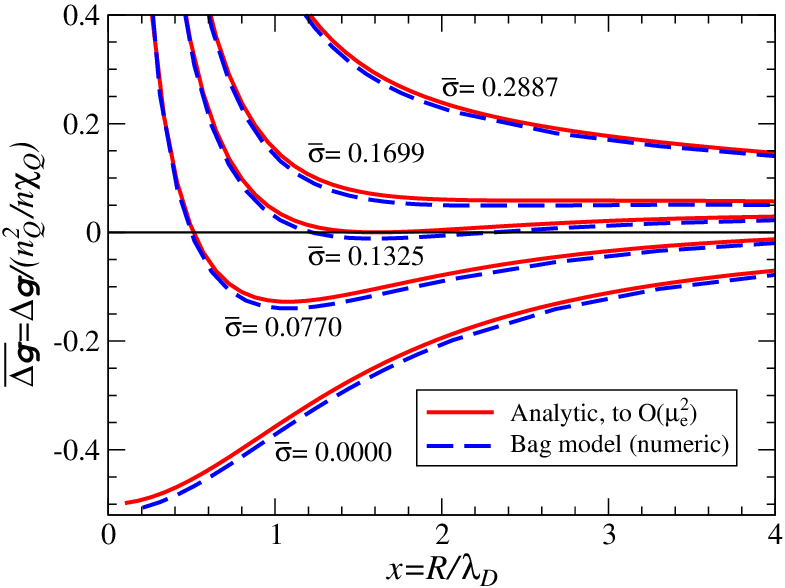}
\caption{The Gibbs free energy 
(i.e. energy per quark) relative to neutral uniform quark matter with $p=0$
for strangelets of various sizes and surface tensions.
We plot $\dGbar(x)$, which is $\De g$ in units of 
$n_Q^2/(n \chi_Q)$, as a function of the radius
$x=R/\lambda_D$, for various values of the surface
tension ($\bar\si$ is $\sigma$ in units of $n_Q^2\lambda_D/\chi_Q$).
The solid lines (red online) are obtained
from (\ref{Delta_g_final}); the dashed lines (blue online) are obtained
from a numerical solution of the Poisson equation, including the
electrons and not making any approximations in the evaluation of $E(R)$.
The numerical results confirm the validity of the approximations made
in deriving (\ref{Delta_g_final},\ref{Delta_g_dimless}).
(The numerical analysis is for a bag model of unpaired quark matter with 
$\mu_{\rm crit}=305$~MeV and $m_s=200$~MeV. The $n_Q$ and $\chi_Q$ for
this model are given in Section \ref{sec:model-dependent}.)
At the critical surface tension $\sibarcrit=0.1325$ 
one can just barely construct a strangelet that 
is favored over uniform quark matter. The radius of this critical strangelet
is $R_* = 1.606\, \lambda_D$. 
For $\sibar<\sibarcrit$, large strangelets and bulk  neutral quark
matter are unstable to fragmentation.  
For $\sibar>\sibar_{\rm no-barrier}=0.1699$, there is no fusion barrier:
strangelets of any size can lower their
free energy by fusing should they encounter each other.
For $\sibarcrit < \sibar < \sibar_{\rm no-barrier}$, there are ``metastable'' droplets, whose
free energy is greater than that of  bulk quark matter but which must overcome
a barrier in order to fuse.
}
\label{fig:EoverN}
\end{figure}

We now evaluate the Gibbs free energy per quark of a strangelet of
radius $R$ relative to that of bulk neutral quark matter, namely
$\Delta g(R)$ of (\ref{Delta_g}).  Integrating the energy density from
(\ref{energy_density}) and adding the surface contribution, $E(R)$ is
given by
\beq
\ba{rcl}
E(R)&=&\dsp\int_0^R d^3 r \left[-p(r) + \mu n 
  - \half q(r) \mu_e(r) \right]+4\pi R^2\sigma\\[2ex]
&=&\dsp \int_0^R d^3 r \left[ -p_0 + \mu n 
  + \half n_Q \mu_e(r) \right] + 4\pi R^2\sigma\ ,
\ea
\label{EofR}
\eeq
where we have used (\ref{charges}) and  (\ref{generic_EoS}) to replace $p(r)$ and $q(r)$.
We see now that the only property of the profile $\mu_e(r)$ 
that we need in order to evaluate $E(R)$ is
the volume average
\beq
\frac{3}{R^3}\int_0^R \!\!d^3r \,\mu_e(r)
  = f \frac{n_Q}{\chi_Q} 
  = \frac{n_Q}{\chi_Q} \left(1-3\frac{x-\tanh x}{x^3}\right) 
\label{volumeavg}
\eeq
where $f$ is the parameter we introduced in making the estimates
(\ref{Delta_g_estimate}) 
and (\ref{crude_estimate}), and which we have now evaluated using the profile $\mu_e(r)$
in (\ref{mueprofile}). 
Thus, by solving the Poisson equation we
have learned that 
\beq
1-f=3\frac{x-\tanh x}{x^3}\ .
\eeq
From (\ref{EofR}) and (\ref{volumeavg}) we find
 \beq
 \frac{E(R)}{N(R)} = \left(-\frac{p_0}{n} +\mu\right) 
  +\frac{n_Q^2f}{2n\chi_Q}+\frac{3\sigma}{nR}
\label{EoverN}
 \eeq
and hence, using (\ref{p0expression}) and (\ref{Delta_g}),
\beq
n \Delta g(R) = \left(\frac{n_Q^2}{2\chi_Q}-p_0\right) 
  +n\left(\mu-\mu_{\rm crit}\right)
- \frac{n_Q^2 (1-f)}{2\chi_Q}+\frac{3\sigma}{R}\ .
\label{Delta_g_semifinal}
\eeq
Notice, however, that the first term is zero when evaluated at
$\mu=\mu_{\rm crit}$ and is given by $n(\mu_{\rm crit}-\mu)$ to lowest
order in $(\mu-\mu_{\rm crit})$. Neglecting fractional errors of
order $(\mu-\mu_{\rm crit})/\mu_{\rm crit}$ in $\Delta g$
(which Fig.~\ref{fig:EoverN} shows are negligible)
we obtain
\beq
n \Delta g(R) = 
- \frac{3 n_Q^2}{2\chi_Q}\frac{x -\tanh x }{x^3}+\frac{3\sigma}{R}\ ,
\label{Delta_g_final}
\eeq
which is (\ref{Delta_g_estimate}) with $(1-f)$ now known but
R* still to be determined.
(By comparing to solutions
obtained numerically, we have confirmed that at large values of
$\sigma/R$ the fractional error introduced in $\Delta g$ is
$\sim\sigma/(nR\mu_{\rm crit})$, as in (\ref{mu_internal}).)
It is convenient to write \eqn{Delta_g_final}
in terms of a dimensionless function 
$\dGbar(x)$ of the dimensionless radius $x=R/\lambda_D$,
\beq
\dGbar(x) =\dsp -\frac{3}{2}\frac{x - \tanh x}{x^3} 
  + \frac{3\sibar}{x}\ ,
\label{Delta_g_dimless}
\eeq
where
\beq
\Delta g(R) \equiv \frac{n_Q^2}{n\chi_Q} \dGbar(x),\qquad
\sibar \equiv \dsp \frac{\chi_Q\si}{\la_D n_Q^2} \ .
\eeq

In Fig.~\ref{fig:EoverN} we plot
$\Delta g$ versus $R$ in dimensionless units
for several values of $\sigma$; this is equivalent to plotting
$\dGbar(x)$ for several values of $\sibar$.
The function $\dGbar(x)$ has a stationary zero at $x_*=1.606$ for
$\sibarcrit=0.1325$:
this corresponds to the marginally stable strangelet at the critical
surface tension.
For $\sibar<\sibarcrit$, the minimum of $\dGbar$ 
occurs at smaller $x$,
and the value of $\dGbar$ at the minimum is negative,
corresponding to a smaller stable strangelet. 
For $\sibar<\sibarcrit$, neutral bulk quark matter and large strangelets can lower
their free energy by fissioning, although for large enough strangelets there
is always an energy barrier to fission.
In the marginally stable droplet with $\sibar=\sibarcrit$ and $x=x_*$, 
\eqn{mu_internal} simplifies to $\mu=\mu_{\rm crit}$, with the
latter two terms in this equation cancelling.  This means that the
quark chemical potential in such a droplet is the same as that
in bulk neutral quark matter.  In a droplet with size $x=x_*$,
the
volume average of $\mu_e$ is reduced from its value
in neutral bulk quark matter by a factor 
$f=0.5051$. 
(Note that we plotted Fig.~\ref{fig:profile}
for $\sibar=\sibarcrit$ and $x=x_*$.)
The parameters in the estimate
(\ref{crude_estimate}) are now fully determined.

We see in Fig.~\ref{fig:EoverN} that at $\sibar_{\rm no-barrier}=0.1699$,
the minimum in $\dGbar(x)$
becomes a stationary inflection point at $x=2.772$. 
For $\sibarcrit<\sibar<\sibar_{\rm no-barrier}$, then, there
exist metastable strangelets with sizes ranging from
$R=1.606\, \lambda_D$ to $R=2.772\, \lambda_D$.
These are stable
against fragmentation, and if two of them encounter each other there
is an energy barrier to their fusion. However, they do have higher
energy per quark than neutral bulk quark matter.
For $\sibar>\sibar_{\rm no-barrier}$
there is no local minimum, and all strangelets can lower their
free energy by fusing with other strangelets should they encounter them.

Converting back to dimensionful quantities, we find 
\beq
\ba{rcl}
\sicrit &=&\dsp
%\frac{n_Q^2\lambda_D}{2\chi_Q}\frac{x_*-\tanh x_*}{x_*^2}\nonumber\\ [2ex]
%&=&\dsp 
0.1325 \,\frac{n_Q^2 \lambda_D}{\chi_Q}
=0.1325\,\frac{n_Q^2}{\sqrt{4\pi\alpha}\chi_Q^{3/2}} \\[2ex]
\si_{\rm no-barrier} &=&\dsp 0.1699 \,\frac{n_Q^2 \lambda_D}{\chi_Q} \ .
\ea
\label{sicrit_result}
\eeq

If the strange matter hypothesis holds (e.g. if $\mu_{\rm
crit}<310$~MeV) and if $\sigma$ in QCD takes on any value less than
$\si_{\rm no-barrier}$, there will be a favored size for the
strangelets that could be found in cosmic rays.  If $\sigma<\sicrit$,
if any strangelets were found they would {\it all} have sizes peaked
around a single value.
If $\sigma$ is somewhat larger, in the regime where there are
metastable strangelets, the distribution of strangelet size in the
universe would include a peak at the size corresponding to the
metastable strangelets, and a continuous distribution of larger
strangelets, big enough that they sit beyond the local maximum in the
$\Delta g(R)$ curve, where $\Delta g(R)$ is a decreasing function of
$R$.  There would be no energy barrier for these larger strangelets to
fuse with one another.

\section{Evaluating the stability of strangelets made from unpaired and 2SC strange quark matter}
\label{sec:model-dependent}

To this point, we have not needed to specify a model for the quark matter
equation of state. Given the values of $n_Q$ and $\chi_Q$ in
any such model (or, ultimately, in the
equation of state of QCD) our results can be used to evaluate the 
critical surface tension $\sicrit$.  If $\sigma < \sicrit$ in QCD, strange
stars will have a solid crust formed from positive strangelets
immersed in an electron gas. And, large strangelets will be unstable
to fission. 

In this section, we provide two bag model examples to illustrate how
our model-independent results translate into estimates of $\sicrit$ in
MeV$/$fm$^2$.  In bag models one writes $p_{\rm QM}(\mu,\mu_e) =
p_{\rm quarks}(\mu,\mu_e) - B$ where $B$ is the bag constant, and
$p_{\rm quarks}$ is obtained by making assumptions, such as treating
the quarks in the bag as noninteracting (``unpaired quark matter''),
or as being in a color superconducting phase, or perhaps as having
some weak residual QCD interactions, although we shall not include the
last effect here.  We begin with a bag model in which the quarks
inside each bag are noninteracting, and then discuss the consequences
of BCS pairing, which would lead to color superconductivity in
infinite quark matter.  We shall set the up and down quark masses to
zero, and treat the strange quark mass $m_s$ as a parameter.  One
extension of our work would be to evaluate $n_Q$ and $\chi_Q$ in
models in which the (strange) quark mass(es) are solved for
self-consistently~\cite{Steiner:2002gx,Abuki:2004zk,Ruster:2005jc,Steiner:2005jm},
rather than taken as a parameter.

\subsection{Unpaired quark matter}

A derivation of the pressure in the bag model for unpaired, noninteracting,
quark matter can be found, for example, in Ref.~\cite{Alford:2002kj}.  
Expanding in  powers of $\mu_e$ according to \eqn{generic_EoS} yields
\bea
p_0(\mu,m_s) &=& -B + \frac{\mu^4}{2 \pi^2}
  + \frac{1}{8\pi^2}\left(2 \mu^3-5\mu m_s^3\right) 
\sqrt{\mu^2-m_s^2}\nonumber\\
&~& + \frac{3 m_s^4}{8\pi^2}\log\frac{\mu+\sqrt{\mu^2-m_s^2}}{m_s}
\label{p0unpaired}\\
n_Q(\mu,m_s) &=& \frac{1}{3\pi^2}\left(\mu^3  
  - \left(\mu^2-m_s^2\right)^{3/2}\right)\sim\frac{m_s^2\mu}{2\pi^2}
\label{nQunpaired}\\
\chi_Q(\mu,m_s) &=& \frac{1}{3\pi^2}\left( 
  5 \mu^2 + \mu \sqrt{\mu^2-m_s^2} \right) \sim \frac{2\mu^2}{\pi^2}
\label{chiQunpaired}
\eea
In the second expressions for $n_Q$ and $\chi_Q$, we have further
assumed that $m_s\ll \mu$.  The resulting simplified expressions yield
results
that are familiar in the literature, for example the fact that
$\mu_e^{\rm neutral}=n_Q/\chi_Q$ gives the well-known 
lowest-order result $\mu_e^{\rm neutral}=m_s^2/(4\mu)$ 
for unpaired quark matter.
We shall not actually use these approximate expressions
in the following, because with $\mu\sim 300$~MeV as in a strangelet or at
the surface of a strange star we cannot necessarily
assume that $m_s\ll\mu$.

Instead of fixing the bag constant $B$ and varying other parameters,
we shall fix the more physical quantity $\mu_{\rm crit}$.
At each value of $\mu_{\rm crit}$ and $m_s$ 
we use \eqn{p0expression} to fix $B$.

\begin{figure}
\includegraphics[width=\hsize]{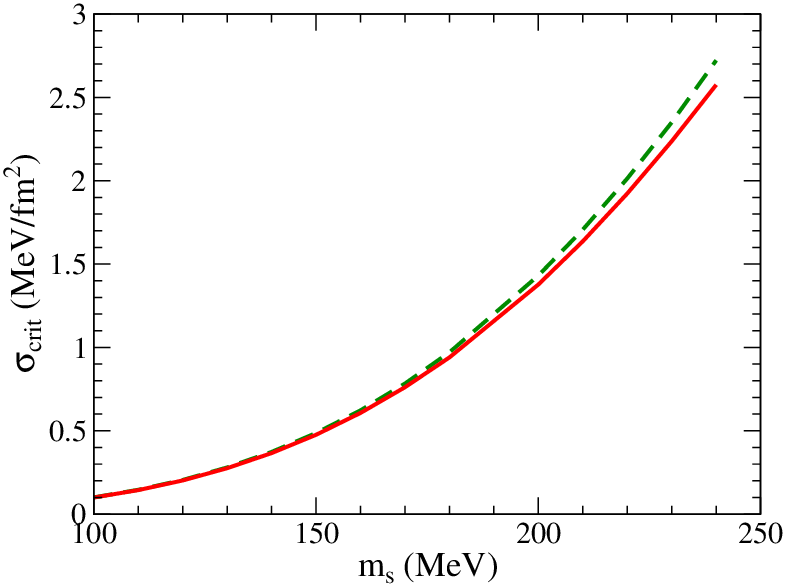}
\caption{The critical surface tension for strangelets made of unpaired
quark matter, as a function of the strange
quark mass, at $\mucrit=305~\MeV$. Other allowed values of $\mucrit$ give
almost indistinguishable curves.  
The solid curve is given by \eqn{sicrit_result}
whereas the dashed curve was obtained by solving the Poisson equation
numerically.
}
\label{fig:sigma_critical}
\end{figure}

In Fig.~\ref{fig:sigma_critical} we show how $\sicrit$
varies with $m_s$ in the bag model for unpaired quark matter
for $\mucrit=305~\MeV$. (The curves for the other allowed
values of $\mucrit$, which vary from $283~\MeV$ to $310~\MeV$,
are indistinguishable from this one except that, as we
discuss below, the $m_s$ at which they terminate is $\mucrit$-dependent.) 
We see that the maximum value of the
critical surface tension is less than $2.7~\MeV/\fm^2$.
The solid curve in the figure is given by
\eqn{sicrit_result} with $n_Q$ and $\chi_Q$ now specified
for the unpaired quark matter bag model
by \eqn{nQunpaired} and \eqn{chiQunpaired}.
The dashed curve was obtained by solving the Poisson equation
numerically, including the electrons, and then evaluating $\Delta g$
without making the approximations that went into deriving 
(\ref{Delta_g_final}).

The curves in Fig.~\ref{fig:sigma_critical} end at $m_s=240$~MeV
because beyond this value nuclear matter becomes unstable
relative to two-flavor quark matter. Unlike the conversion to
three-flavor quark matter, which must surmount a barrier of many weak
interactions, this would allow rapid conversion of nuclei to quark
matter. For $\mu_{\rm crit}=305$~MeV, requiring that the pressure of
two-flavor quark matter at $\mu=310$~MeV be negative
requires $m_s \leqslant 240$~MeV.  Note that we can
consider values of $m_s$ all the way up to $\mu$, at the expense of
tuning $\mu_{\rm crit}$ closer and closer to $\mu_{\rm
nuclear}=310$~MeV as we take $m_s\rightarrow\mu$ in such a way as to
keep nuclear matter (just) stable with respect to two-flavor quark
matter.  In this fine-tuned limit, we can get strange quark matter
with an arbitrarily small strange quark density, with $n_Q$ and
$\chi_Q$ arbitrarily close to their two-flavor values $\mu^3/(3\pi^2)$
and $5\mu^2/(3\pi^2)$ respectively.  From \eqn{sicrit_result} we see
that in this limit, $\sicrit\rightarrow 0.01185\, \mu_{\rm nuclear}^3
/ \sqrt{12\pi^3 \alpha}=5.502$~MeV$/$fm$^2$.  This is the absolute
limit to how large $\sicrit$ can be pushed in the bag model for
unpaired quark matter.
Fig.~\ref{fig:sigma_critical} shows the limit
if one requires $\mu_{\rm crit}$ to be 5 MeV below $\mu_{\rm
nuclear}$.

% At ms=100 B is 144 to 159,  mu_crit is 283 to 310
% At ms=200 it is 144 to 150, mu_crit is 297 to 310
% At ms=250 it is 144 to 147  mu_crit is 304 to 310

To give a sense of the scales that characterize the critical
strangelets, let us take noninteracting quark matter with $\mu_{\rm
crit}=305$~MeV and $m_s=200$~MeV as an example.  We find
$n_Q=0.07104$~fm$^{-3}$, $\chi_Q=0.4644$~fm$^{-2}$, and
$\lambda_D=4.845$~fm.  Bulk neutral quark matter has $\mu_e^{\rm
neutral}=n_Q/\chi_Q=30.1$~MeV and has quark number density
$n=0.9093$~fm$^{-3}$.  If $\sigma=\sicrit=1.377$~MeV, the critical
strangelets with radii $1.606\lambda_D=7.782$~fm have baryon number
$A\approx 598$ and charge $Z\approx 69$, consistent with the result
$Z\approx 0.1 A$ from the literature~\cite{Madsen:1994vp}.  If $A$
were significantly smaller, as for example would be the case for
$\sigma$ significantly below $\sicrit$, it would be important to
include curvature energy and shell effects in the calculation of
$E/N$.  The analysis of Ref.~\cite{Madsen:1994vp} indicates that these
can reasonably be neglected for strangelets with $A$'s greater than a
few hundred.

Finally, let us discuss the effect of interactions. Already in the
earliest bag model analyses of strange stars and
strangelets~\cite{Farhi:1984qu}, the perturbative QCD interactions
between quarks inside a bag were taken into account.  To zeroth order
in $m_s$, these just introduce modifications to 
the relationship between $\mu$ and $n$ which are small
if $\al_s$ is small. We leave the inclusion of such perturbative effects
to future work, and turn to the possibility of color superconducting
quark matter, which can introduce changes to the values of
$n_Q$ and $\chi_Q$ that are qualitative, in the sense that they
do not become arbitrarily small  when some parameter of the EoS is
varied.

\subsection{Color superconducting quark matter}

The strong interaction between quarks is attractive in the
color-antisymmetric channel, and this leads to BCS pairing and color
superconductivity in cold quark matter~\cite{Reviews}.  The critical
temperature for color superconductivity is expected to be in the range
of tens of MeV, which is far above the temperature of strangelets or
neutron stars. We therefore work at $T=0$ throughout.

Unlike perturbative interactions, color superconductivity can have
dramatic qualitative consequences for the properties of quark matter,
because those quarks which undergo BCS pairing have their number
densities ``locked'' into being equal, as the pairing energy gained in
so doing overcompensates for the free energy cost of maintaining
number densities that would not minimize the free-particle free
energy. In this section we explore the consequences of color
superconductivity for the stability of strangelets.

To sketch the context for our discussion, let us describe
the neutral color-superconducting phases that may
occur in quark matter with $\mu\approx 310~\MeV$, depending
on the strength of the attractive quark-quark coupling that leads to BCS pairing. 
We expect the color-flavor locked
(CFL) phase at the strongest values of the coupling, with 
some other, probably non-isotropic, phase occurring at intermediate coupling,
and unpaired quark matter at lower couplings still.  As we will see, the
non-isotropic phases are expected to yield stable strangelets similar
to those arising from unpaired quark matter, whereas 
the analysis of the stability of CFL quark matter
strangelets is qualitatively different.  We will see that the
two-flavor-paired ``2SC'' phase is not expected to be the favored
state of neutral quark matter at $\mu\approx 310~\MeV$, but it may
well play a role in strangelets, which consist of charged quark
matter.

The CFL phase \cite{Alford:1998mk,Reviews}
is the ground state of three-flavor quark matter when the attractive quark-quark
coupling is sufficiently strong that its gap parameter satisfies
$\Delta_{\rm CFL}>m_s^2/(2\mu)$~\cite{Alford:2003fq,Alford:2004hz}.
In the CFL phase, quarks of all three colors and all three flavors undergo BCS
pairing~\cite{Alford:1998mk}, and the resultant locking of the
Fermi surfaces makes it an electromagnetic insulator, 
neutral in the absence of 
electrons~\cite{Rajagopal:2000ff,Alford:2002kj}, with
$n_Q=\chi_Q=0$ and $\mu_e^{\rm neutral}=n_Q/\chi_Q=0$.
If bulk quark matter at $\mu_{\rm crit}$ is in the CFL
phase, strange stars will be neutral with $\mu_e=0$ in their
interiors~\cite{Usov:2004iz}, meaning that there is no possibility of
charge separation and fragmentation, no possibility of a solid crust,
and no reason for large strangelets to fission into smaller ones.  This can
be seen from our analysis by noting that in the CFL phase $\sicrit=0$.

At intermediate values of the quark-quark
coupling, we know that a different color superconducting 
phase must occur.
For $m_s^2/(5.2 \mu) \lesssim \Delta_{\rm CFL} < m_s^2/(2\mu)$,
model analyses that are restricted to isotropic phases
predict a gapless CFL (gCFL) phase~\cite{Alford:2003fq,Alford:2004hz}.
Depending on the value of $\Delta_{\rm CFL} \mu/m_s^2$, 
the values of $n_Q$, $\chi_Q$ and $\mu_{\rm neutral}$  in gCFL quark matter can
fall anywhere between those characterizing the CFL and
unpaired phases, and in particular can be arbitrarily small but nonzero.
This would make the investigation of strangelets in some gCFL-like
phase very interesting, as the Debye length could be arbitrarily large. 
However, the gCFL phase per se 
is unstable to the formation of current-carrying
condensates~\cite{Casalbuoni:2004tb,Huang:2004bg,Kryjevski:2005qq,Fukushima:2006su,Gerhold:2006dt} and so it cannot be the ground state.
The nature of the ground state at intermediate coupling is not yet established, but
one possibility is a three-flavor crystalline color superconducting
phase~\cite{Casalbuoni:2005zp,Ciminale:2006sm,Mannarelli:2006fy,Gerhold:2006dt},
with a nontrivial crystal structure like that favored in the simpler two-flavor 
case~\cite{Alford:2000ze,Bowers:2002xr}.
Such phases do {\em not} involve the locking of Fermi surfaces:
this is in large part why they are well-motivated candidates to
replace the gapless CFL phase. This means that, unlike in the CFL or gCFL phases,
their values of $n_Q$, $\chi_Q$ and hence $n_Q/\chi_Q$ are 
likely similar to those of
unpaired three-flavor quark
matter~\cite{Casalbuoni:2005zp,Mannarelli:2006fy}, so that strangelets
made of such matter will have similar stability properties to ones
made of unpaired quark matter.

The final possible color superconducting phase we consider is the
``2SC phase'' in which BCS pairing occurs only between up and down
quarks of two
colors~\cite{Bailin:1983bm,Alford:1997zt,Rapp:1997zu,Reviews}.  
Previous studies indicate that we should
not expect this phase to occur in bulk {\em neutral} quark matter 
near the transition to the vacuum (which we are assuming
occurs at $\mu_{\rm crit}<310$~MeV). In bag models in which $m_s$ is
a parameter,
the neutrality constraint ensures that 
either unpaired, CFL or gCFL quark matter is always favored over 2SC at $T=0$
\cite{Alford:2002kj,Alford:2004hz}. In models that solve for $m_s$ 
self-consistently there is sometimes a small 2SC window,
but only if the interaction and parameters are chosen such that $m_s$ 
is much too large for three-flavor quark matter to exist with 
$\mu=\mu_{\rm crit}<310$~MeV~\cite{Steiner:2002gx,Abuki:2004zk,Ruster:2005jc,Steiner:2005jm}.

However, even if  the 2SC phase is not favored in bulk matter, which
must be neutral, it may occur
in the charged quark matter of the
stable (or metastable) strangelets that we have constructed.  We
see in Fig.~\ref{fig:profile}
that $\mu_e$ inside these droplets is significantly less than
$\mu_e^{\rm neutral}$.  This means that the unpaired $u$ and $d$ Fermi
surfaces are closer to each other, and farther from the $s$ Fermi surface,
than in neutral unpaired bulk matter.  This should favor the 2SC phase.
To investigate this, we take results from Ref.~\cite{Alford:2002kj} 
and use them to evaluate the parameters of the generic quark matter
EoS \eqn{generic_EoS},
\bea
p_0^{\rm 2SC}(\mu,m_s) &=& 
  p_0^{\rm unpaired} +\frac{\Delta_{\rm 2SC}^2\mu^2}{\pi^2}
\label{p02SC}\\
n_Q^{\rm 2SC}(\mu,m_s) &=& 
  n_Q^{\rm unpaired} +\frac{\Delta_{\rm 2SC}^2\mu}{3\pi^2}
\label{nQ2SC}\\
\chi_Q^{\rm 2SC}(\mu,m_s) &=& 
 \frac{1}{18\pi^2}\left( 12 \mu^2 + 6 \mu \sqrt{\mu^2-m_s^2}+\Delta_{\rm 2SC}^2
 \right)\nonumber\\
&\sim& \frac{\mu^2}{\pi^2}\ .
\label{chiQ2SC}
\eea
(These expressions have all been derived assuming 
$\Delta_{\rm 2SC}\ll \mu$, which implies
$\Delta_{\rm 2SC}=2^{1/3} \Delta_{\rm CFL}$~\cite{Reviews}.)  Note the
change in $\chi_Q$: it is approximately half as large as in unpaired
quark matter, meaning that the Debye length in 2SC quark matter is
larger than that in unpaired quark matter by a factor of $\sqrt{2}$.
This change once again originates in the locking of those
Fermi surfaces which pair, so it is qualitative, in the
sense that it is independent of the value of $\Delta_{\rm 2SC}$.
To see how big an effect this could have on the stability
of strangelets, imagine giving the unpaired phase a larger bag constant than
the 2SC phase, so that bulk neutral quark matter is in the 2SC phase.
The results of section \ref{sec:construction} show that the
size of the critical strangelet would be increased by a factor of
approximately $\sqrt{2}$, and $\sicrit$ would be increased by a factor of
approximately $2\sqrt{2}$.
However, if we keep the same bag constant in unpaired and 2SC quark matter
then the effect is smaller. For example, let us
analyze the case where
$m_s=200$~MeV, $\Delta_{\rm 2SC}=25$~MeV and $\mu_{\rm crit}^{\rm
unpaired}=305$~MeV.  We find $n_Q^{\rm 2SC}=1.012\, n_Q^{\rm
unpaired}$ and $\chi_Q^{\rm 2SC} = 0.479\, \chi_Q^{\rm unpaired}$.  
In this case we cannot naively use the results of section 
\ref{sec:construction} (yielding $\sicrit$ enhanced by the factor of about
$2\sqrt{2}$ described above) because
$\mu_{\rm crit}^{\rm 2SC}$ is $305.507$~MeV, larger than
$\mu_{\rm crit}^{\rm unpaired}$, and consequently the critical
strangelet occurs when the $\Delta g(R)$ curve constructed for the 2SC
phase has a local minimum with $\Delta g(R)=\mu_{\rm crit}^{\rm
unpaired}-\mu_{\rm crit}^{\rm 2SC}<0$, corresponding to 2SC
strangelets with the same $E/N$ as bulk neutral unpaired quark matter.
We then find $\sicrit=2.79~\MeV/\fm^2$ and $R_*=1.16\,\lambda_D^{\rm
2SC}=8.1$~fm. Comparing this to the results for unpaired quark
matter strangelets at the same $\mucrit$ and $m_s$ (which gave
$\sicrit=1.38~\MeV/\fm^2$ and $R_*=7.78~\fm$),
we see that the occurrence of 2SC matter has changed $\sicrit$ by
a factor of about 2. The increase in $\la_D$ has been cancelled by a decrease
in $x_*$ with the result that $R_*$ hardly changes. Evaluating $A$ and $Z$
for the critically stable 2SC droplet, we find $A\approx 677$ and $Z\approx 105$.

It should be noted
that the 2SC Cooper pairs, whose size is $\sim
1/\Delta_{\rm 2SC}$, fit within these strangelets, but not by
much.  This means that for a specified interaction strength between
quarks, the parameter $\Delta_{\rm 2SC}$ occurring in our analysis,
which describes the effect of $ud$-pairing on the $E/N$ of our
strangelets, will not have exactly the same value as that describing
2SC pairing in bulk matter.  Had we taken $\Delta_{\rm 2SC}$
significantly smaller, our analysis of 2SC pairing would not even be
qualitatively reliable.

We conclude that if the attraction between quarks is strong
enough that the critical strangelets are in the 2SC phase (but not so
strong as to favor the CFL phase for
bulk neutral quark matter) then $\sicrit$ is
increased by a factor which could be as large as $\approx 2\sqrt{2}$
if $\Delta_{\rm 2SC}$ were tuned such that $\mu_{\rm crit}^{\rm
2SC}\rightarrow \mu_{\rm crit}^{\rm unpaired}$, but which is more
typically smaller, of order 2.

\section{Discussion}
\label{sec:discussion}

If the surface tension in QCD of an interface between quark matter and
vacuum is below a critical value $\sicrit$, and if the strange quark
matter hypothesis holds, astrophysicists may observe (or may be
observing) strange quark stars that have crusts made of positively
charged strangelets, with size of order the Debye length $\lambda_D$,
immersed in an electron gas. And, larger strangelets will be unstable
to fission.  In Sections \ref{sec:stability} and
\ref{sec:construction}, we have developed a model-independent
evaluation of $\sicrit$, along the way constructing profiles of
strangelets taking into account both the energy benefit of charge
separation and the energy costs introduced by the surface tension and
Debye screening.  Our result is given in \eqn{sicrit_result}, in terms
of two parameters $n_Q$ (the charge density of quark matter with
$\mu_e=0$) and $\chi_Q$ (the charge susceptibility of quark matter
with $\mu_e=0$) which are not currently known from first principles.
Given any equation of state, whether from a model 
or from a future full QCD calculation, 
$n_Q$ and $\chi_Q$ can be evaluated in terms of the quark
number chemical potential $\mu$ and the strange quark mass $m_s$.

Our purpose in Section \ref{sec:model-dependent} was to get a sense of
the size of $\sicrit$, in MeV$/$fm$^2$, using as a guide bag models
for unpaired quark matter and quark matter in the 2SC color
superconducting phase.  A new twist in this section was the
possibility that neutral bulk quark matter and stable strangelets
could be in different phases (unpaired and 2SC respectively).  We
conclude from the model-dependent investigation in this section that
it is easy to find parameters for which $\sicrit\sim
1-3$~MeV$/$fm$^2$. And, by fine-tuning both $m_s$ and $\Delta_{\rm
2SC}$, it is possible to push $\sicrit$ to as large as
$5-7$~MeV$/$fm$^2$.  Although these conclusions are model-dependent,
our results phrased in terms of $n_Q$ and $\chi_Q$ as in Section
\ref{sec:construction} can be applied in future to any model, for any
phase, as our understanding of dense quark matter continues to
improve.

The surface tension $\sigma$ is not known in QCD, but it has been
calculated in the bag model for unpaired quark
matter~\cite{Berger:1986ps}.  It ranges from about $10$~MeV$/$fm$^2$
for $m_s=100-120$~ MeV to about $4~\MeV/\fm^2$ for $m_s=250$~MeV.
If the interface has a thickness of order 1 fm, rather than being
infinitely sharp as assumed in the bag model, the surface tension is
almost certainly larger.
It seems most likely that $\si>\sicrit$,
making large strangelets stable against fission and giving strange
quark stars, if they exist, fluid surfaces as in
Ref.~\cite{Alcock:1986hz}.  However, the combination of the bag model
estimate for $\sigma$ from Ref.~\cite{Berger:1986ps} and our
model-dependent investigation in Section \ref{sec:model-dependent}
leave open the possibility of the opposite conclusion.

Further
investigation of the properties of the strange star crusts that result from 
fragmentation and charge separation at strange star surfaces, as proposed
in Ref.~\cite{Jaikumar:2005ne}, is warranted.  We have only analyzed the
surface of the crust, where $p_{\rm ext}=0$ and strangelets are spherical.
If $\sigma$ is less than $\sicrit$, we must then analyze the deeper layers
of the crust.
Preliminary numerical
work indicates that $\Delta g$ becomes less negative with increasing
$p_{\rm ext}$, but to date this ignores the alternative geometrical shapes of
the strangelets that will occur below the surface of the crust. Furthermore, in
the inner half of the crust, characterized by electron-filled ``voids''
embedded in quark matter
rather than by strangelets embedded in an electron-gas,
the curvature energy which we have neglected must be included.  
The curvature energy is positive for strangelets, acting, like the
surface tension, to suppress small strangelets.
For voids, however, the curvature
energy is negative, acting in opposition to the surface tension.  
If charge-separated crusts do occur on strange stars,
it seems likely that they will prove more similar
to conventional neutron star crusts than to the fluid surface previously
considered for strange stars.  It remains to be investigated, however,
{\em how} similar to neutron star crusts they prove to be,
and in particular whether the rich X-ray burst phenomenology
observed in accreting compact stars can be consistent with a
strange star crust.

It remains unlikely that strange star crusts can give rise to
pulsar glitches, since glitches require a crust within with
charged nuggets (nuclei or strangelets would be equivalent) are
immersed in both an electron gas and a superfluid. However,
there is no superfluid in the strange star crust proposed in
Ref.~\cite{Jaikumar:2005ne}.
However, if quark matter in the
crystalline color superconducting phase occurs somewhere within the
core of a strange star, this could be the locus in which pulsar
glitches originate~\cite{Alford:2000ze,Bowers:2002xr}.

Another way in which observations could settle some of the
questions we have raised here would be the discovery of strangelets
in cosmic rays, with a peak in the distribution of $A$ and $Z$
corresponding to the baryon number 
$A_{\rm stable}$ of the most favored stable (or metastable)
strangelet.  If $\sigma<\sicrit$, and if strangelets occur in cosmic
rays, any strangelets whose initial baryon number
(in whatever astrophysical
collision produced them) was much larger than $A_{\rm stable}$
would fission into strangelets with 
$A \gtrsim A_{\rm stable}$.  
Strangelets with $A$ only slightly bigger than $A_{\rm stable}$
will not fission because of the high energy energy per quark
of the smallest strangelet produced.
We therefore expect a peak in the strangelet distribution
at $A\approx A_{\rm stable}$, tailing off to no strangelets at all
over some range of larger values. 
If the estimates of
Ref.~\cite{Madsen:2004vw} of the flux of strangelets in cosmic rays
incident on the earth are correct, the single-strangelet detection
capability of the AMS-02 detector, scheduled to go into operation on
the International Space Station in a few years, extends to strangelets
with all $A<10^4$~\cite{AMS}.  For two particular choices of model and
model parameters in Section \ref{sec:model-dependent}, we found 
critical strangelets with $A\approx 600$ and $Z\approx 70$ in one  case
and $A\approx 680$ and $Z\approx 105$ in the other case. More
generally, we expect that if $\sigma<\sicrit$ the stable strangelets
will have values of $A$ lying between many hundred and a few
thousand. If quark matter has such a small surface tension, then
both strangelets and a peak in their size-distribution should be
within the discovery reach of AMS-02.

\vspace{0.8cm}

\noindent
{\bf Acknowledgements:} 
%We thank various people for useful discussions. 
KR acknowledges the
hospitality of the Nuclear Theory Group at LBNL. This research was
supported in part by the Offices of Nuclear Physics and High
Energy Physics of the Office of
Science of the U.S.~Department of Energy under contracts
\#DE-FG02-91ER50628, \#DE-FG01-04ER0225 (OJI), W-7405-ENG-36 and
\#DE-AC02-05CH11231 and cooperative research agreement
\#DF-FC02-94ER40818.

%\nocite{*}
%\bibliography{ms}
%\bibliographystyle{apsrev}

\end{document}